\documentclass[useAMS,usenatbib]{mn2e}
\usepackage{psfig}
\usepackage{amsmath}
\usepackage{amssymb}
\def \cov {\text{cov}}
\def \mgii {Mg\,{\sc ii}}
\def \civ {C\,{\sc iv}}
\def \hbeta {H$\beta$}
\def \empha {({\emph a})}
\def \emphb {({\emph b})}
\def \emphc {({\emph c})}
\def \emphd {({\emph d})}

\title[Composite reverberation mapping]{Composite reverberation mapping}

\author[Fine et al.]
       {S. Fine$^1$\thanks{stephen.fine@durham.ac.uk},
         T. Shanks$^1$, S. M. Croom$^2$, P. Green$^3$, B. C. Kelly$^3$, E. Berger$^3$, \newauthor
R. Chornock$^3$, W. S. Burgett$^4$, E. A. Magnier$^4$, P. A. Price$^5$ \\
$^1$Department of Physics, Durham University, South Road, Durham DH1
         3LE, UK \\
$^2$Sydney Institute for Astronomy, School of Physics, The University of Sydney, NSW 2006, Australia \\
$^3$Harvard-Smithsonian Center for Astrophysics, 60 Garden Street, Cambridge, MA 02138, USA \\
$^4$Institute for Astronomy, University of Hawaii at Manoa, Honolulu, HI 96822, USA \\
$^5$Department of Astrophysical Sciences, Princeton University, Princeton, NJ 08544, USA\\
}

\begin{document}

\maketitle

\begin{abstract}

Reverberation mapping offers one of the best techniques for studying
the inner regions of QSOs. It is based on cross-correlating continuum
and emission-line light curves. New time-resolved optical surveys will
produce well sampled light curves for many thousands of QSOs. We
explore the potential of stacking samples to produce composite
cross-correlations for groups of objects that have well sampled
continuum light curves, but only a few ($\sim2$) emission-line
measurements. This technique exploits current and future wide-field optical monitoring surveys (e.g. Pan-STARRS, LSST) and the multiplexing capability of multi-object spectrographs (e.g. 2dF, Hectospec) to significantly reduce the observational expense of reverberation mapping, in particular at high redshift (0.5 to 2.5).

We demonstrate the technique using simulated QSO light curves and
explore the biases involved when stacking cross-correlations in some simplified situations. We show
that stacked cross correlations have smaller amplitude peaks compared
to well sampled correlation functions as the mean flux of the emission
light curve is poorly constrained. However, the position of the
peak remains intact. We find there can be `kinks' in stacked
correlation functions due to different measurements contributing to
different parts of the correlation function. While the magnitude of the kinks must be fitted for, their positions and relative strengths are known from the spectroscopic sampling distribution of the QSOs making the bias a one-parameter effect. We also find the S/N in the correlation functions for the stacked and well-sampled cases are comparable for the same number of continuum and emission line measurement pairs.

Using the Pan-STARRS Medium-Deep Survey (MDS) as a template we show that cross-correlation lags should be measurable in a sample size of 500 QSOs that have weekly photometric monitoring and two spectroscopic observations. Finally we apply the technique to a small sample (42) of QSOs that have light curves
from the MDS. We find no indication of a peak in the stacked cross-correlation. A larger spectroscopic sample is required to produce robust reverberation lags.


\end{abstract}


\section{Introduction}

%
%
%
%
%

The inner regions of active galactic nuclei (AGN) offer a unique
opportunity to study matter
within a few parsecs of a super-massive black hole. Reverberation
mapping is designed to study (primarily) the broad-line region (BLR) of AGN by
measuring the interaction between continuum and broad-line flux
variations \citep{b+m82,pet93}. The physical model assumes the BLR is
photoionised by a UV continuum that is emitted from a much smaller
radius. Variations in the ionising continuum produce equivalent
variations in the broad emission line flux after a delay that can be associated with the light travel time.

Reverberation mapping of a single system requires many epochs of
emission-line and continuum luminosity measures. A peak in the cross
correlation between the two light curves indicates the time lag
between continuum and emission-line variations. To date lags have been
measured for some tens of objects following this approach
\citep{pea04,ben06,ben10,den06,den10}.

Reverberation mapping has led to significant advances in the
understanding of AGN (e.g. the radius--luminosity relation;
\citealt{wpm99,kas00}, stratification and kinematics of the BLR
\citealt{p+w99,p+w00}, black hole mass estimates \citealt{pea04}
etc.). However, these campaigns
are observationally expensive and time consuming as they require many
observations of individual objects over a long period of time.

Rather than focusing on obtaining many hundreds of epochs of
observations on single objects this paper explores an observational
technique that is only now becoming possible due to the new generation
of time-resolved photometric surveys and the multiplexing capabilities
of multiple-object spectrographs (MOS).

Time resolved surveys, such as those being performed with the
Pan-STARRS1 (PS1) telescope \citep{kai02}, will measure the broad-band
(continuum) light curves of many thousand of QSOs. The PS1 medium-deep survey
(MDS) is taking images of ten 7\,deg$^2$ fields every few days in
five photometric bands. Number counts imply there are $\sim500$ QSOs
with $g<22$ in each of these fields \citep{croom09b} making them
easily surveyable 
with current MOSs in $\sim$a night of observing time.
Repeating the spectroscopic observations regularly would
allow traditional reverberation mapping to be performed on samples of
thousands of QSOs. However, this would require a large amount of observing
time on highly subscribed telescopes to produce results. In this paper
we look at the potential for reverberation mapping in stacked samples
of QSOs that only have a few ($\sim2$) spectroscopic epochs of data,
but are coupled with well-sampled continuum light curves.

In section~\ref{sec:ooa} we outline the principle of stacking cross
correlations, then in section~\ref{sec:sims} we simulate QSO light
curves to illustrate the technique, in section~\ref{sec:ps_sim} we present an empirical simulation of a QSO survey and the potential results, in section~\ref{sec:rdata} we
apply the technique to early data from a spectroscopic survey of QSOs
in the MDS region and in section~\ref{sec:sum} we summarise the
results of our investigation. Throughout this paper we use a flat
$(\Omega_{\rm m},\Omega_{\Lambda})=(0.3,0.7)$, $H_{0}=70\,{\rm
  km\,s}^{-1}\,{\rm Mpc}^{-1}$ cosmology.

\section{Outline of approach} 
\label{sec:ooa}


Observed light curves are made up of a series of (often unevenly sampled) discrete measurements. Estimates of the cross-correlation between continuum and emission line light curves (denoted $C$ and $L$ below) are therefore limited by their sampling.
A variety of techniques have been developed for estimating cross correlations with observational samples
\citep{g+s86,g+p87,z+p11}. In this paper we 
will focus on the discrete cross covariance function \citep{e+k88} as
it lends itself simply to the stacking technique. The discrete cross
covariance is calculated in terms of pairs of observations
($C_i,L_j$). Taking all
pairs of continuum and emission line observations such
that the time lag $t_i-t_j$ is between $\tau$ and $\tau+\delta\tau$,
then the cross covariance amplitude for lag $\tau$ is estimated with
\begin{equation}
X(\tau)=\displaystyle\sum\limits^{t_i-t_j\in[\tau,\tau+\delta\tau]}_{\substack{i,j}}
\frac{(C_i-\overline{C})(L_j-\overline{L})}{n_{pair}}
\label{equ:dxc}
\end{equation}
where $n_C$ and $n_L$ are the number of continuum and emission line measurements respectively. $\delta\tau$ defines the bin size in the cross-correlation. Here we will use fixed bin sizes, but see \citet{ale97} for the use of variable bin sizes to optimise results.
Equation~\ref{equ:dxc} is the same as that for the cross correlation except
that it is not normalised by the rms of each variable. $\delta\tau$ defines the bin size in the cross-correlation. 
In this paper we will focus on the covariance rather than the correlation function as they are almost equivalent in terms of the analysis presented
here but using the cross covariance clarifies the discussion in later
sections.

The discrete cross covariance is a function on pairs of
observations. Assuming all measurements have the same errors and well defined
mean levels the variance of the cross covariance is inversely
proportional to the number of data-data pairs. Therefore, desirable
results require a large number of pairs of
observations at a wide range of time lags. Traditionally this is
achieved through repeat observations of an object
over a period of years to build up sufficient data at all time lags.
However, there is no explicit reason that all of the data must come
from a single source. Given a group of objects that have similar
variability properties we are able to combine the data-data
pairs from each object in the group to obtain a `composite' cross covariance function with higher S/N than is obtainable for the individual objects. Essentially this process is the same as calculating the covariance function for each object and then stacking them.

This stacking technique has been used before for continuum auto covariance/correlation analysis \citep{alm00,vdb04,wil07}, but to our knowledge it has not previously been applied to cross covariances and/or reverberation mapping. 






\section{Simulating data}
\label{sec:sims}

%
%



To demonstrate the principle of stacking cross covariances we
simulate QSO continuum and emission line light curves.
We model the continuum as a first-order auto-regressive process
following (e.g. \citealt{kel09,mac10,zu12}). Simulating this process requires a
damping timescale $\tau_c$ and amplitude of short-time-scale
variations $\sigma_c$ that we fix for each simulation.

The emission line flux at a given
time is defined by convolving the continuum light curve with a given
transfer function that is a function of $\tau$
(e.g. \citealt{e+k88}). The transfer function is not well
constrained observationally for AGN. For simplicity we use a Gaussian with
unit area centred on $\tau=\tau_l$ with full-width-half-maximum
$\tau_l/2$. Note that this is not a physically motivated transfer
function, but is roughly similar to empirical transfer functions for
Balmer lines that have been reverberation mapped \citep{hwp91,ben10}.

After simulating the light curves we add random Gaussian
noise into the continuum and emission line data with rms $e_c$
and $e_l$ to simulate uncorrelated effects and measurement error.

In total there are five input parameters. For each set of simulations
we perform we will create many light curves. Each time we simulate a
light curve we draw the input parameters
from Gaussian distributions with a mean and rms fixed for that set of
simulations and table~\ref{tab:in_pars} gives a list of these parameters.

\begin{table}
\begin{center}
\caption{Parameters used when simulating light curves (see text for
  the definitions of each parameter). In each case the parameter used
  in a simulation is drawn from a Gaussian distribution with mean and
  rms given in this table. Parameters that are required to be
  positive have their distribution functions truncated at zero.}
\label{tab:in_pars}
\begin{tabular}{cccccc}
\hline \hline
Parameter & $\sigma_C$ & $\tau_c$ & $\tau_l$ & $e_C$ & $e_L$ \\
units & flux & days & days & flux & flux \\
\hline
mean & 5. & 20. & 40. & 0.2 & 0.2 \\
rms  & 1. & 5.  & 5., 30.  & 0.1 & 0.1 \\
\hline \hline
\end{tabular}
\end{center}
\end{table}

\subsection{Stacked vs. non-Stacked cross covariance functions}

\begin{figure}
\centerline{\psfig{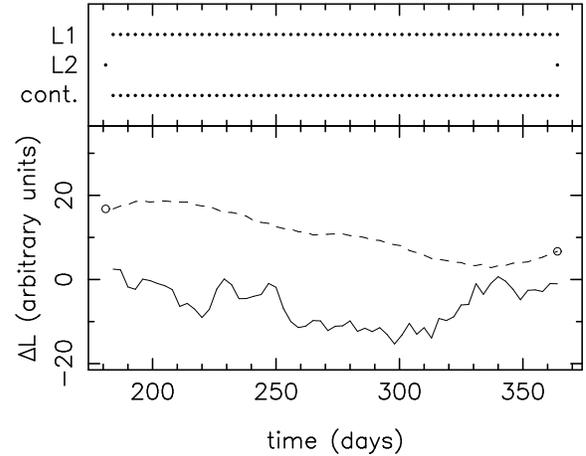}}
\caption{(\emph{bottom}) Example of a simulated continuum (solid) and emission line
  (dashed) light curve. The dashed curve has been offset from the
  solid for clarity. (\emph{top}) The dots in the top panel show how the
  continuum and emission line are sampled in the case of L1 and L2.}
\label{fig:lc_eg}
\end{figure}

To demonstrate the principle of stacking cross covariances we will
concentrate on two simplified scenarios.
In each we assume a QSO is observed photometrically (giving us
continuum luminosity measures) every three days for six months of the
year. In the first
case (L1) we assume that at every date we get photometric data we also
obtain a spectrum and emission-line flux measure. In the second case
(L2) we only obtain line
fluxes twice a year at either end of the continuum sampling. The L1 scenario is designed to be equivalent to standard reverberation mapping with large numbers of emission line and continuum measurements, L2 demonstrates the stacking technique. We
illustrate these situations in Fig.~\ref{fig:lc_eg}. The top panel in
the figure shows the sampling over six months of the continuum and
emission line in each of the cases. Below we show a simulated
light curve of the continuum (solid line) and the emission line (dashed
line) over the period (solid line) using the parameters from
table~\ref{tab:in_pars}.

%
%

\begin{figure*}
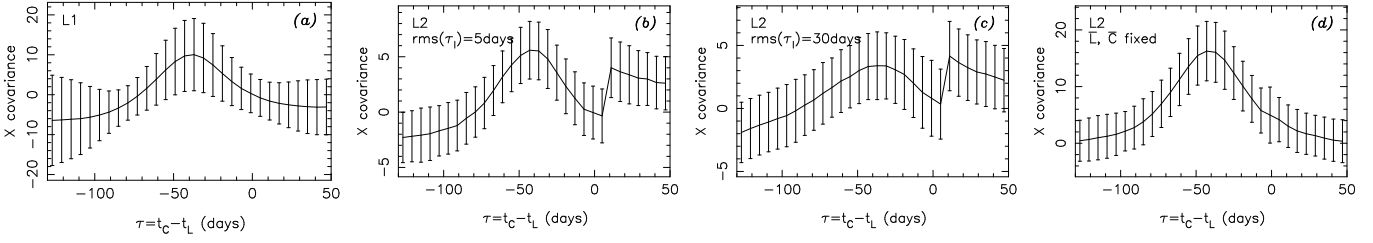

\centerline{\psfig{file=plot_xc_nL61_nq1.ps,width=4.5cm,angle=-90}
\hspace{-0.2cm}
\psfig{file=plot_xc_nL2_nq30.ps,width=4.5cm,angle=-90}
\hspace{-0.2cm}
\psfig{file=plot_xc_nL2_nq30_sig30.ps,width=4.5cm,angle=-90}
\hspace{-0.2cm}
\psfig{file=plot_xc_nL2_nq30_Fav.ps,width=4.5cm,angle=-90}}
\caption{Comparison between the mean cross covariance functions for a
  single QSO with 61 epochs of spectroscopic data (L1; \emph{a}) and 30 QSOs
  with only two epochs (L2; \emph{b}). \emphc\ shows the same simulations as \emphb\ except the distribution of lags used is 40$\pm$30\,days rather than $\pm$5. In \emphd\ the L2 simulation is re-run with $rms(\tau_l)=\pm5.$
  but the cross covariance functions are calculated using the known
  values of $\overline{L}$ and $\overline{C}$ rather than calculating
  them from the simulated observations.}
\label{fig:xc_comp1}
\end{figure*}

If we take just a single year's worth of data there are 61 continuum
measurements and 61 and 2 emission line measurements for L1 and L2
respectively. In the case of L1 we create 1000 simulated continuum and
emission line light curves. We calculate the cross covariance
function in each case and in Fig.~\ref{fig:xc_comp1}\empha\ we plot the
average of the 1000 covariance functions along with the rms between
simulations as error bars.

Since there are $\sim30$ times the number of spectroscopic
measurements in L1 we compare this with a case where 30 QSOs have been
sampled as in L2. The cross covariances from the 30 QSOs are then
stacked to produce an ensemble covariance function. We simulate this
situation 1000 times and plot the results in Fig.~\ref{fig:xc_comp1}\emphb\ and \emphc, the difference between these simulations is that the first has the rms of $\tau_l=5.$\,days, while the second has $\tau_l=30.$\,days to illustrate the effect of increasing the scatter in reverberation lags.

It is apparent that, in our simulations at least, we can reproduce the
peak in the cross covariance function at $\tau_l=40$ using stacked
covariance functions. However, there are differences between the cross
covariances for L1 and L2. In the following subsections we discuss
the most prominent of these.



\subsubsection{Peak height in stacked vs. non-stacked cross
  covariances}
\label{sec:pbias}

The peak in the covariance function for L2
is biased low because we are under sampling the emission line light
curve and so are unable to define the mean level precisely. The value
of $\overline{L}$ in equation~\ref{equ:dxc} is heavily biased by the
two individual measurements of $L_i$, hence the covariance function
is weaker. We can derive the expected level of this bias in the case
of $n_L$ measurements of the emission line luminosity. Taking the
definition of the discrete cross covariance:
\begin{eqnarray}
X(\tau) &=& \sum \frac{(C_i-\overline{C})(L_j-\frac{\sum
    L_k}{n_L})}{n_{pair}} \nonumber \\
&=& \sum \frac{(C_i-\overline{C})(L_j-\frac{L_j}{n_L} - \frac{\sum
    \limits_{k\neq j}L_k}{n_L})}{n_{pair}} \nonumber \\
&=& \sum \frac{\frac{n_L-1}{n_L}(C_i-\overline{C})(L_j -
    \overline{L}_{k\neq j})}{n_{pair}}
\label{equ:pbias}
\end{eqnarray}
(note the sums here are over the same data points as equation~\ref{equ:dxc})
$\overline{L}$ in equation~\ref{equ:pbias} is now calculated over the
$n_L-1$ emission line measurements not including $L_j$ and so is
unbiased by $L_j$. Hence the amplitude of a cross covariance calculated
from $n_L$ emission line measurements will be proportional to
$(n_L-1)/n_L$. This is $\sim$a factor of two in the case of L1
compared to L2 in our simulations as is illustrated in
Fig.~\ref{fig:xc_comp1}.

Note that had we calculated cross correlation rather than covariances
this effect is not as important since the rms of the $L_i$s (that
normalises a cross correlation) is affected in the same manner as the
covariance.

\subsubsection{Errors in stacked vs. non-stacked cross covariances}

The rms error bars on the L1 cross covariance are not all
equal and decrease from left to right. In the case of L2 the error
bars are roughly constant. The different ways that the emission line
light curves are sampled for L1 and L2 mean that the number of
data-data pairs contribution to the covariance function at a given
$\tau$ is not the same in each case.

\begin{figure}
\centerline{\psfig{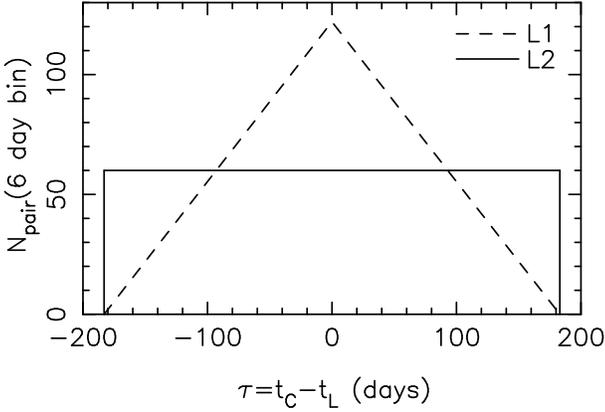}}
\caption{The number of emission-line and continuum data--data pairs
  that are used in the cross-covariance calculation. We compare the L1
  and L2 examples (see text) with a bin size of six days.}
\label{fig:comp_12}
\end{figure}

In Fig.~\ref{fig:comp_12} we plot the number of data-data pairs
contributing to a covariance function as a function of
$\tau$ for L1 and L2. The constant time sampling of L1 leads to
a more concentrated distribution of lags.
In the case of L2 the distribution is flat while the
total number of pairs is approximately the same for L1 and L2. This
leads to an increase in the noise of L1 cross covariances in the
under-sampled areas of lag space around six months.

In general the errors are considerably larger in the case of L1
compared to L2. This is because L1 over samples the light curve we have
simulated and, while the number of measurements going into the L1 and
the stacked L2 cross covariances is roughly the same, the emission-line
measurements are correlated in the case of L1, increasing the
noise.

\subsubsection{The `kink' in the L2 cross covariance}

The `kink' in the L2 cross covariance occurs because we have only two
measurements of the emission line luminosity, and because the mean of
the continuum level is not accurately defined.

At any lag in an
individual cross covariance for L2 (before stacking) there is only
one data-data pair available to define the cross covariance. All of
the positive lags use the first line measurement, and all of
the negative lags use the second. The kink in
Fig.~\ref{fig:xc_comp1}\emphb\ occurs as we move from one regime to
the other. To explain the kink we take the
definition of the discrete cross covariance in the case of just two
line measurements:
\begin{equation}
X(\tau) = \sum \frac{(C_i-\overline{C})(L_j-\overline{L})}{n_{pair}} \nonumber
\end{equation}
\begin{equation}
= (C_i-\overline{C})(L_j-(L_1+L_2)/2)
\label{equ:kink1}
\end{equation}
since there are only two emission line measurements and only one
continuum---line pair contributes at a given lag. If we use
$L_{j\prime}$ to denote the emission line measurement that isn't $L_j$
then
\begin{equation}
X(\tau) = 0.5(C_i-\overline{C})(L_j-L_{j\prime}).
\label{equ:kink2}
\end{equation}
We now calculate the expectation value of the cross covariance.
We use angled brackets to denote this expectation value that is
averaged over a large number of realisations, this is distinct from
the average of the individual continuum luminosities ($\overline{C}$)
that is calculated over the number of observations of a single object.
\begin{equation}
\langle X(\tau) \rangle = 0.5(\langle C_i L_j \rangle - 
\langle C_i L_{j\prime} \rangle - \langle \overline{C} L_j \rangle +
\langle \overline{C} L_{j\prime} \rangle) \nonumber 
\end{equation}
we notate the covariance of two variables as $\cov(x,y)=\langle xy
\rangle -  \langle x \rangle  \langle y \rangle$
\begin{equation}
\begin{array}{c}
\langle X(\tau) \rangle = 0.5(\langle C_i \rangle \langle L_j \rangle + \cov(C_i,L_j) -
\langle C_i \rangle \langle L_{j\prime} \rangle - \cov(C_i,L_{j\prime}) \\
- \langle \overline{C} \rangle \langle L_j \rangle -
\cov(\overline{C},L_j) +
\langle \overline{C} \rangle \langle L_{j\prime} \rangle + 
\cov(\overline{C},L_{j\prime}))  \nonumber 
\end{array}
\end{equation}
\begin{equation}
= 0.5(\cov(C_i,L_j) - \cov(C_i,L_{j\prime}) - \cov(\overline{C},L_j) + 
\cov(\overline{C},L_{j\prime})).
\label{equ:kink3}
\end{equation}
The first term of equation~\ref{equ:kink3} is the quantity we are
trying to measure with the cross covariance. The other terms are
biases. The factor of $0.5$ is due to there only being two emission
line measurements (see section~\ref{sec:pbias}). For the second term
note that $\cov(C_i,L_1)<\cov(C_i,L_2)$. However, this term will be
small with respect to the other biases. For the other terms:
$\cov(\overline{C},L_1)$ will be small as $\overline{C}$ is
calculated from the continuum light curve that is measured after $L_1$
while $L_1$ is dependent on the continuum level before it is measured. 
$\cov(\overline{C},L_2)$ on the other hand is significant. This is
partly due to the manner in which we have constructed our
simulation. The auto-regressive continuum light curve means
that continuum points are covariant with their
neighbours, as we may expect is the case in reality. Furthermore, the transfer function that is used to define
the emission line luminosity makes $L_2$ covariant with a number of
the continuum points.

Given the values in Table~\ref{tab:in_pars} we can calculate
$\cov(\overline{C},L_1)$, $\cov(\overline{C},L_2)$, $\cov(C_1,L_2)$
and $\cov(C_{n_C},L_1)$ to get the expected offset in the cross
covariance between $\tau<0$ and $\tau>0$. These are respectively
0.43, 5.04, 0.03, 0.0004 in the units used. Hence we expect the offset
to be $\sim4.6$ in Fig.~\ref{fig:xc_comp1}\emphb. This is in good agreement with our simulations.

In Fig.~\ref{fig:xc_comp1}\emphd\ we reproduce the simulations in \emphb, except that rather than calculating $\overline{C}$ and $\overline{L}$ from the simulated light curves, we use their correct values as defined in the simulation. In this case the bias is not apparent and there is no kink in the covariance function.


\subsection{Optimising the stacked results}

The major source of bias in the stacked covariance functions is the poorly defined mean levels, particularly $\overline{L}$. The precision of $\overline{C}$ can be improved simply with more measurements and, in the practical example we give below of the PAN-STARRS survey there will eventually be many more than the $\sim$60 photometric measurements that we have assumed here.

While more spectroscopic measurements would increase the precision of $\overline{L}$ this somewhat goes against the principle of the technique. Less biased values for $\overline{L}$ can be estimated from $\overline{C}$ and the global equivalent width distribution at fixed luminosity. However, this would also include a significant loss of precision and would smooth the kink in the correlation function at the expense of S/N. Since the time sampling of the continuum and emission line is known, the location of the kink and the shape (if not magnitude) of its effect on the covariance function can be estimated. Removing the bias from the results then requires a one-parameter fit to the data. In this case the magnitude of the bias can hold important information on the covariances between emission line and continuum emission in QSOs and, accurately measured, could help studies of transfer functions and the interactions between the accretion disk and BLR.

The distribution of the reverberation lags of the quasars that are stacked in this manner has the effect of smoothing out the stacked covariance function. This broadens the peak in the stacked covariance function and hence reduces the signal in the peak. The two values of $rms(\tau_l)$ used in the above simulations are used to illustrate this effect with extreme values. The lack of large numbers of objects that have several reverberation mapped lines means that we cannot be certain of the distribution of lags for high redshift quasars. However, the \hbeta\ line has been mapped for a significant number of Seyferts. While there has been a considerable range of lags measured, almost all of this variation has been shown to be due to the radius--luminosity relation. \citet{kasp05} find only $\sim15$\,\%\ intrinsic scatter around this relation. Furthermore, the small degree of scatter in \mgii\ and \civ\ line widths for the brightest quasars may indicate that there is even less intrinsic scatter in the radius--luminosity relation in that regime \citep{me2,me3}. Hence it would be advisable to use quasars with a small range of luminosities when calculating a stacked cross covariance to improove the signal. Indeed this may be advisable anyway since one of the primary applications of this technique would be to evaluate the radius-luminosity relation for UV quasar lines.

To get a better idea of what results we may expect from a feasible survey of quasars we design a simulation based on what is currently known about AGN. In the next section we present this simulation to give a realistic impression of the results that could be achieved with this technique.

\section{A physically motivated simulation for stacked reverberation mapping with the Pan-STARRS MDS}
\label{sec:ps_sim}

In the remainder of this paper we make a preliminary application of the method in the Pan-STARRS medium deep survey (MDS).
The MDS consists of ten $7$\,deg$^2$ fields that are imaged every few
nights in five photometric bands ($g_{P1}r_{P1}i_{P1}z_{P1}y_{P1}$). The size of the fields,
along with QSO number counts, make each field surveyable with the
current generation of multi-object spectrographs in $\sim$one night of
observing time, yielding $\sim$500 QSOs (with $g_{P1}<22$). The regular
photometric monitoring of the fields means that for every spectrum
taken there are many continuum-emission line data pairs that can be
used in cross covariance analyses.

To evaluate the potential for reverberation mapping in the MDS fields we run simulations based on current knowledge of QSO variability parameters.
We assume that a QSO survey is performed similar to the 2SLAQ QSO survey \citep{croom09b} and simulate a single MDS field of data with the following prescription.
\begin{itemize}
\item{From 2SLAQ number counts we assume 500 QSOs total. We therefore draw 500 QSOs at random from the 2SLAQ catalogue but only accept those objects at the right redshift to have \mgii\ in an optical spectrum ($0.4<z<2.4$).}
\item{For each object we use the BH mass estimates in \citet{me2} and scaling relations from \citet{mac10} to obtain the continuum variability parameters $\sigma_C$ and $\tau_C$ in the observed $g_{P1}$-band assuming 0.3 and 0.15\,dex scatter about the mean relations respectively.}
\item{We extrapolate from the $i$-band magnitude to 5100\,\AA\ assuming a power-law continuum of $f_\lambda\propto\lambda^{-1.5}$ and use the radius-luminosity relation from \citet{kasp05} with 0.1\,dex scatter to obtain the lag for the emission line response $\tau_L$ under the assumption that the \mgii\ and H$\beta$ lines have similar lags \citep{m+j02}.}
\item{Timescales are converted to the observed frame and continuum light curves are simulated assuming 6\,day sampling over an eight month period. Emission line flux values are calculated as previously with one spectrum at the beginning and end of the continuum points.}
\item{We then add random scatter to the continuum points based on the error on the 2SLAQ magnitude and add 10\,\%\ scatter to the emission line points to simulate measurement errors.}
\end{itemize}

\begin{figure}
\centerline{\psfig{file=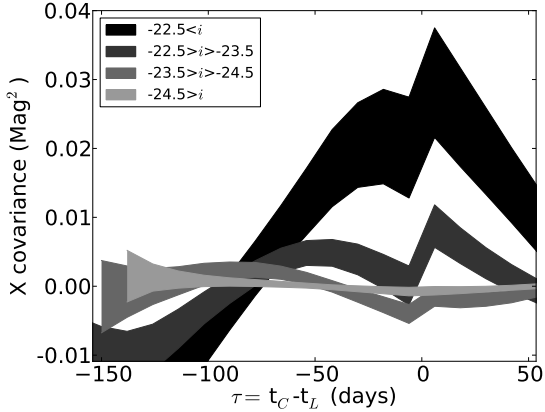,width=8cm}}
\caption{Simulated cross covariance functions between \mgii\ emission-line and $g_{P1}$-band light curves for a survey of an MDS field. We assumed there would be 500 objects in the field and simulated individual QSO parameters by drawing randomly from the 2SLAQ catalogue. The objects are split into four bins by absolute magnitude to show the tendency towards longer lags in brighter objects.}
\label{fig:mds_sim}
\end{figure}

In Fig.~\ref{fig:mds_sim} we bin eight months worth of data by the $i_{P1}$-band absolute magnitude and plot the stacked cross-covariances. The shaded areas show the regions within the rms of the mean cross-correlation when using 12\,day bins in the discrete cross-covariance. The figure shows that the tendency towards longer lags in brighter QSOs may be detectable in a single year's worth of data on a single MDS field given the above assumptions.

\begin{figure}
\centerline{\psfig{file=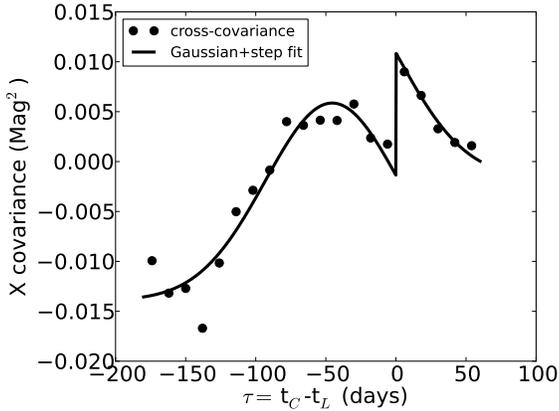,width=8cm}}
\caption{Simulated cross-covariance function for a single magnitude bin in one of our simulations (points) and the functional fit to it (solid line). Here a Gaussian fits the simulated points and the kink at $\tau=0$\,days is corrected by the step function.}
\label{fig:mds_sim_ind}
\end{figure}

For each individual realisation in our simulations we try to find the peak in the cross-covariance functions. We fit a Gaussian, plus an offset, plus a step at $\tau=0$\,days. The step function corrects for the kink in the covariance functions at $\tau=0$\,days. In Fig.~\ref{fig:mds_sim_ind} we show the simulated cross-covariance for a single magnitude bin in one of our realisations along with the fitted Gaussian$+$step.

\begin{figure}
\centerline{\psfig{file=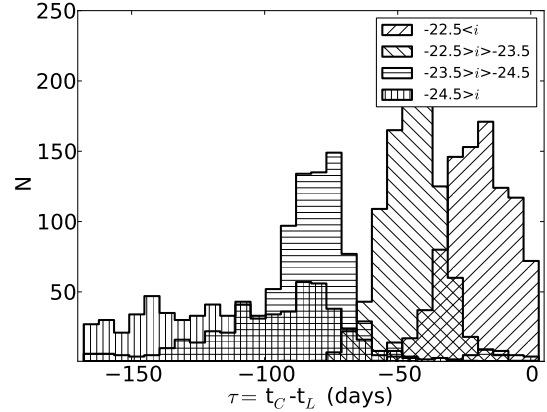,width=8cm}}
\caption{The distribution of the peaks in individual simulated cross-covariances, binned by $i_{P1}$-band magnitude, as defined by Gaussian fits. the tendency for longer lags in the brighter bins is clear here. The brightest bin has lags that are too long to be properly sampled and we only find a good peak in $\sim40$\,\%\ of the simulations.}
\label{fig:mds_sim_hist}
\end{figure}

Using the fit Gaussians we find the location of the peak in the cross correlations. The distributions of these values are shown in Fig.~\ref{fig:mds_sim_hist} for each of the magnitude bins. Note that we do not constrain any of the parameters in the fits but still find a peak in the range of timescales sampled for almost all ($>$99\,\%) of the realisations except in the brightest magnitude bin. In the brightest bin the timescales are longer and we only obtain a lag in $\sim40$\,\%\ of the realisations.
Note that in some cases a Gaussian gives a poor automated fit to the cross-covariance. More careful fitting in individual cases would give better results. The mean $\tau_L$ values in the simulated light curves are 27, 54, 97 and 227\, days for the faintest to brightest magnitude bins. The fact that we measure considerably smaller lags than this is due to the fainter objects in each bin being more variable and hence contributing more to the stacked correlation function.

\subsection{The potential of stacked cross-covariance functions}

We have demonstrated that, at least in our simulations, it is possible to retrieve an average radius--luminosity relation for a sample of quasars. If possible this would be of considerable interest for quasar studies. The radius--luminosity relation has not been strongly constrained for any emission lines other than \hbeta. However, UV lines (notably \mgii\ and \civ) are commonly used to estimate virial black hole masses in high-redshift quasars (e.g. \citealt{m+j02,vest02}). These virial mass estimates are based on the assumption that a radius--luminosity relationship exists for these UV lines equivalent to that measured for \hbeta. An observational determination of these relations, even if only in averaged stacks, would be of great use in determining black hole masses in high-redshift quasars.

The simulations performed in section~\ref{sec:ps_sim} assumed a parent sample of 500 quasars. The MDS survey as a whole would contain $\sim5000$ quasars to the flux limit we assume. It is not beyond current telescopes to survey this number of objects $\sim$twice yearly. Such a dataset, in particular if built up over several years, would offer the potential to make extremely high-precision stacked covariance functions. These could offer a unique opportunity to study the transfer function and hence the structure of the BLR although such studies would have to be mindful that convolved into the covariance function is the distribution of different lags and transfer functions of the different quasars in a stack.

Multiple lines could be mapped for the same stacks of quasars. These data give relative information on the stratification of the BLR and where the various emission zones are in quasars. Combined with dynamical measurements (i.e. line profiles) these give indications of the dominant motions in the BLR.

The potential gains from being able to reverberation-map high-z quasars are significant. Above we have outlined just a few. This paper presents a technique that we believe offers a feasible rout towards obtaining these results.

Recent studies have also suggested either narrow- or broad-band `photometric' reverberation mapping as an observationally cheaper method for obtaining reverberation lags for quasars (e.g. \citep{c+l73,has11,c+d12}). While neither technique has been proven for large-scale samples the relative gains of each technique are somewhat unclear. Narrow-band photometric mapping requires the quasars are at the correct redshift for the filter, and so cannot be applied on such large scales as the other two methods. Broad-band photometric reverberation mapping is more efficient than the technique described here in terms of the spectroscopy required. However, a single epoch of spectroscopy is required to obtain accurate redshifts. The complexity of decoupling continuum and emission line light curves and the effects of having several lines in a single filter requires extremely accurate photometry and complex reduction techniques. On the other hand broad-band photometric reverberation mapping does not require the complexity of obtaining flux-calibrated fibre spectra with high enough S/N to calculate precise line fluxes. Finally, a major gain for stacked covariance functions is the flat sampling distribution (Fig.~\ref{fig:comp_12}). Bright quasars can have continuum timescales of years, and correspondingly large BLR lags. Obtaining good constraints on these lags requires sampling many times the continuum$+$lag timescale in classical reverberation mapping. However, in our technique one needs only sample $1\times$ the continuum$+$lag timescale and then information is built up through stacking.


The problem of obtaining robust reverberation mapping results in the high-redshift Universe is a problem that may be solved with current and planned time-resolved photometric surveys. In the rest of the paper we apply our technique to a small number of objects that already have the correct observations available.

\section{Pan-STARRS MDS early data}
\label{sec:rdata}

In this section we derive MDS light curves for a sample of QSOs that have $>1$ spectra, and calculate the stacked cross-covariance function for the sample.

\subsection{PS1 light curves}

The PS1 telescope \citep{hod04} is performing a series of time-resolved photometric surveys of the northern sky. We are particularly interested in the MDS as offering the best opportunity for our analysis. Images of the ten MDS fields are taken every four-five nights in each of the photometric bands while not affected by the sun or moon. On each night eight dithered exposures are taken and combined to form a nightly stack (see e.g. \citealt{kai10,ton12} for further details of the PS1 telescope, observing strategy and data processing).
We used {\sc psphot}, part of the standard PS1 Image Processing Pipeline system \citep{mag06}, to extract point-spread-function photometry from
nightly stacked images of the MDS fields. Each nightly stack is
divided into $\sim70$ skycells. We calibrate each of these separately
using SDSS photometry \citep{fuk96,yor00} of moderately bright
($16<\text{Mag.}<18.5$) 
point sources (defined as having $type=6$ in the SDSS database). The
bright magnitude cut is used since the brightest objects can create
artifacts in the PS1 images. We use 18.5 as the faint cut so that we
have a large number of objects used in the calibration of each skycell
while ensuring high-precision photometry for each object (95\,\% have
errors$<0.01$\,Mag.).

\begin{figure}
\centerline{\psfig{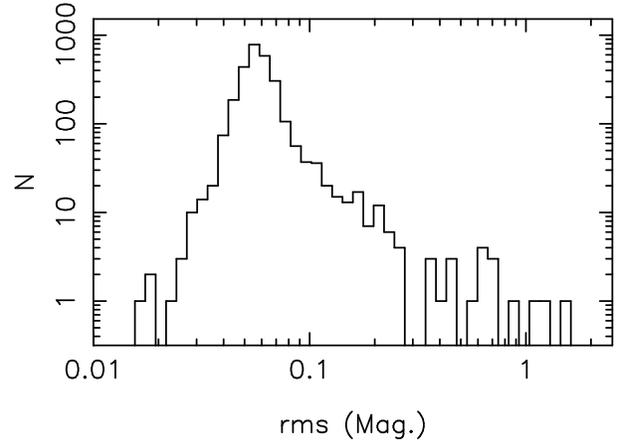}}
\caption{The rms of the calibration offset between PS1 $g_{P1}$ and SDSS $g$-band photometry in the MD03 field. In cases where the rms is $>0.1$\,Mag the calibration is deemed to be suspect and the skycell is rejected. This corresponds to $\sim5$\,\%\ of the individual skycells.}
\label{fig:rms_hist}
\end{figure}

We found that in some individual cases the flux calibration was unstable due
primarily to artifacts around objects used in the flux calibration
in the PS1 imaging. To calibrate our data
we measure an average offset between the PS1 and SDSS photometry,
we also record the rms around this average for each skycell we calibrate.
Fig.~\ref{fig:rms_hist} shows a histogram of the rms of the
calibration offset for each epoch of each skycell in the $g_{P1}$-band for
the MD03 field. The tail to larger rms' is indicative of PS1 skycells
that have poor magnitudes for the stars used in the flux calibration. We only accept skycells with $\text{rms}<0.1$\,Mag. This cut removes $\sim5$\,\%\ of the stacked skycells from our sample.

\subsection{Hectospectra}

Spectra of QSOs in the MDS fields are being taken as part of an
ongoing project to study the variability of QSOs.
QSO candidates are selected for spectroscopy using current
photometric databases of QSOs based on SDSS photometry \citep{ric09,bov11}
in addition to which point sources that correspond to X-ray sources,
variability selected objects and UVX selected objects are targeted
for spectroscopy.

The MDS fields are being surveyed with the Hectospec instrument
on the MMT. Each MDS field is tiled with seven MMT
pointings. Exposures are $\sim$1.5\,h in length meaning that an MDS field
($\sim500$ QSOs) can be surveyed in $\sim1$\,night of on-sky observing
time.

The spectra are extracted and reduced using standard Hectospec
pipelines \citep{min07}. They are then flux calibrated using observations of F
stars in the same fields. These stars are compared with a grid of
model stellar spectra, created using the spectral synthesis code {\sc spectrum} \citep{g+c94,ggh01} with models from \citet{c+k04}, to
correct for the response of the 
Hectospec instrument. Absolute flux calibration is then made by
fitting to the $r$-band SDSS magnitudes of the stars. Errors in the
response correction are typically $\lesssim10$\,\% over the main part
of the spectrum ($\sim$4000--8500\,\AA). However, uncertainty on the absolute flux calibration can have a larger effect on the line fluxes we measure. This depends strongly on the number of stars used in the calibration and is typically $\lesssim10$\,\%. However, in a handful of Hectospec fields some
stars are significantly off the average calibration. In general the spectra of these stars are fainter than expected and may be affected by small positioning
errors. These stars are removed manually from the calibration. However, the same problem may affect QSO spectra in our sample.

\subsection{Our sample}

\begin{figure}
\centerline{\psfig{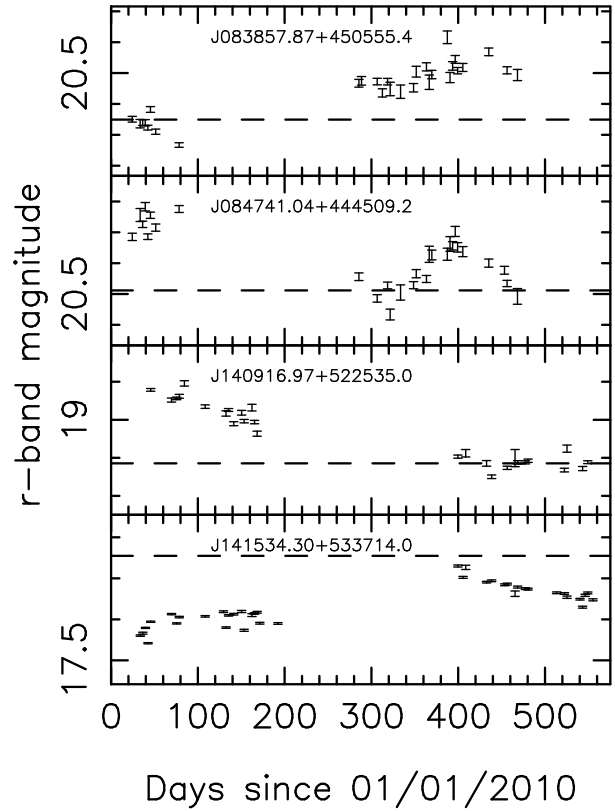}}
\caption{Four example $r_{P1}$-band light curves of QSOs in our sample. The
top two and bottom two come from different MDS fields (MD03 and MD07
respectively). All of the objects show evidence for variability over
the $\sim1$\,year of monitoring. Dashed lines in each plot indicate
the SDSS $r$-band magnitude.}
\label{fig:obs_lc}
\end{figure}

In total spectra of 855 ($g<22$) QSOs have been obtained, primarily in MDS fields MD03 and MD07. More information on the selection of this QSO sample and the spectra will be given in an upcoming paper. Further QSO spectra were kindly taken by the MDS transient team. Most QSOs in the sample have only one
spectrum and so cannot be used to cross correlate with the
continuum observations.
However, 82 of these objects are in the SDSS DR7 QSO catalogue \citep{sch10}. These objects have spectra taken with the Sloan telescope giving us two spectra over a time baseline of $\sim10$\,years (for details of the Sloan spectrograph and SDSS QSO selection see \citealt{gun06,sto02,rich02b}). Furthermore, we have taken $>1$\,spectrum of 59 QSOs in our sample giving us a spectroscopic baseline from $\sim$50 to $\sim$100\,days. In Fig.~\ref{fig:obs_lc} we show the $r_{P1}$-band light curve for 4 QSOs selected from our sample as examples.

\subsection{Emission line fluxes}

\begin{figure}
\centerline{\psfig{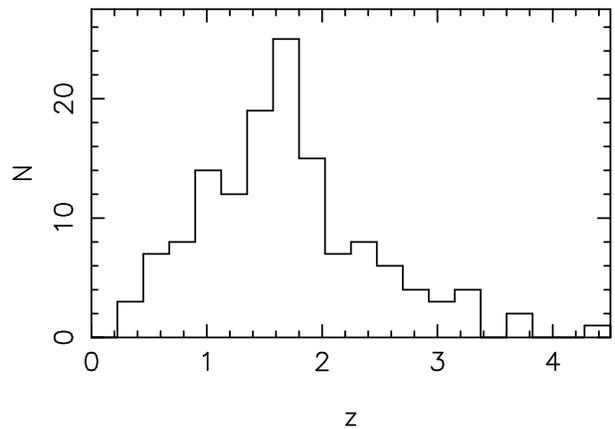}}
\caption{The redshift distribution of the 138 objects that have MDS
  light curves and $>1$ spectroscopic observation from the MMT and
  SDSS. The distribution is typical of optically/UV selected samples
  and is highly incomplete in the interval $2\lesssim z\lesssim 3$.}
\label{fig:nz}
\end{figure}

In Fig.~\ref{fig:nz} we show the redshift distribution of the 138 QSOs
with $>1$\,spectrum. Their distribution is relatively
typical of that of the SDSS (and other optically selected) QSO
catalogue with the vast majority between $z\sim0.5$ and 2. This
corresponds roughly to the range of
redshifts over which the \mgii\ line is redshifted into optical
spectra and we focus on the \mgii\ line for the rest of this work.

We fit the \mgii\ line following the prescription outlined in
\citet{me2} and each fit is manually inspected to check the reliability. We
calculate emission line flux and error directly from the
continuum-subtracted spectrum following \citet{car98}. The typical S/N
of these lines is $\sim3-30$ for the SDSS spectra and $\sim10-200$ for
the Hectospec spectra. Hence in the case of the Hectospec spectra the
error on the line fluxes are dominated by the flux-calibration errors
rather than the spectral S/N. The SDSS spectra, that have a flux
calibration error of $\sim5$\,\% \citep{ade08}, are more typically dominated
by the statistical noise in the spectra.

\begin{figure}
\centerline{\psfig{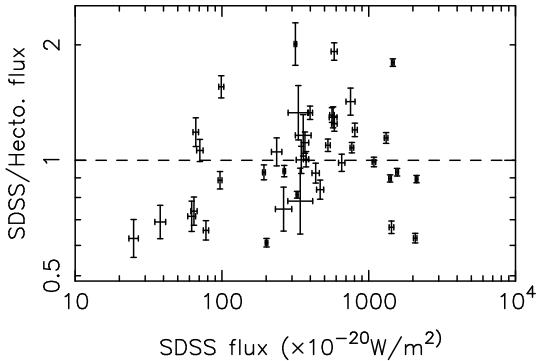}}
\caption{This figure compares the \mgii\ line flux measured from the
  archive SDSS spectra and during the Hectospec survey. Errors on the
  SDSS measurements are dominated by the spectral S/N, while in the
  Hectospec observations they are dominated by uncertainty in the flux
  calibration.}
\label{fig:SHflux}
\end{figure}

Through visual inspection we remove a number of spectra that have
i) broad-absorption lines, ii) unusual spectra that are poorly
characterised by our fitting or iii) are affected by residual sky/telluric
features. In Fig.~\ref{fig:SHflux} we show the measured SDSS and
Hectospec fluxes for the 42 QSOs that had good \mgii\ flux
measurements from each spectrum. These span $\sim$two orders of
magnitude in luminosity and cover the redshift range $0.45<z<1.68$.

\subsection{The binned cross covariance}

\begin{figure}
\centerline{\psfig{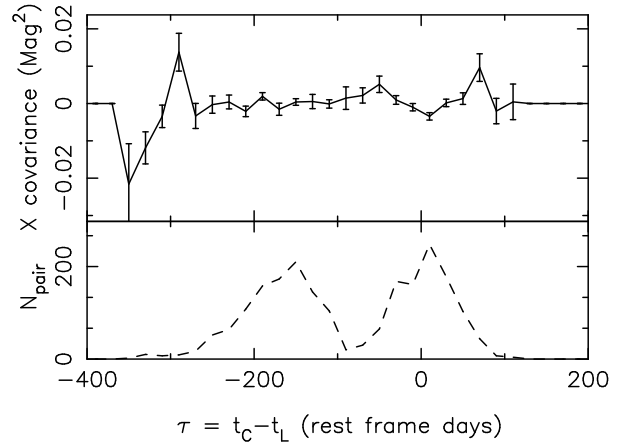}}
\caption{The stacked discrete cross covariance function for the 42
  objects in our final sample. The bottom panel shows the number of
  emission-line and continuum data--data pairs contributing to each
  bin. The bins are 20\,days (rest-frame) in width.}
\label{fig:xc_mg}
\end{figure}

We calculate the discrete covariance function for the 42 objects
with two good \mgii\ flux measurements. In the calculation we convert
the emission-line fluxes to magnitudes, hence we are measuring the
fractional (rather than absolute) cross covariance. We perform the
calculation in bins with width 20\,days in the rest frame of the
observed QSOs. In Fig.~\ref{fig:xc_mg} we show the covariance
function along with the number of data-data pairs in each time-lag
bin. The small number of objects and spectra means that there is
little signal in the cross covariance. Furthermore, due to the
observing times we have a gap in our time sampling at
$\sim90$\,days.

The small number of objects involved means that we were unlikely to find a lag in this data.
Fig.~\ref{fig:xc_mg} is given rather as
an example of what can be done with current data. Note that we do not expect the type of step bias shown in Fig.~\ref{fig:xc_comp1}\emphc\ since, in those objects that have spectra from the SDSS, only one of the two spectra is contributing to the correlation function at these lags.

\section{Summary}
\label{sec:sum}

We have demonstrated a technique for stacking cross-correlations and covariances. We focus on QSO emission line and continuum light curves in the case when only one is well sampled. We demonstrate the technique via a suite of simple simulations that highlight some of its biases and limitations as well as its advantages. While the stacked analyses show smaller peaks and can produce erroneous steps in the results due to the limited time-sampling, the position of the peak remains unbiased. Furthermore, given similar total numbers of emission line and continuum measurements the technique gives comparable S/N compared to the classic, unstacked approach.

We focus on the PAN-STARRS MDS and show that it may be possible to measure average reverberation lags for QSOs in these fields with a relatively small investment of telescope time based on empirical simulations of QSOs in these fields. Finally we performed a stacked cross-covariance analysis on 42 QSOs from the MDS that have well sampled continuum emission and $>2$ spectra from our observations and the SDSS archive. We find no indication of a peak in these data although the small numbers mean this would have been unlikely. In the near future multi-epoch spectroscopic observations of MDS fields would be required to allow for stacked lags to be measured.

We note that the same technique could be applied outside of reverberation mapping. The relationship between X-ray and optical variability is complex in QSOs \citep{she03,mrm08,are09} and lags between X-ray and optical light curves can be used to investigate the dominant emission processes at these wavelengths. Here the highly sampled optical data could be combined with wide-field X-ray imaging data (e.g. XMM) as an efficient means of producing optical to X-ray cross correlations. The gain in efficiency would be proportional to the number of X-ray QSOs that can be simultaneously imaged.

\section{acknowledgements}

The authors would like to acknowledge Peter Draper for assistance in setting up the source extraction software used in this paper. We would also like to thank the MDS transient team for sharing MMT fibres and adding significantly to the size of our QSO sample.

The data in this paper came from the SDSS, Pan-STARRS and the MMT.

Funding for the SDSS and SDSS-II has been provided by the Alfred P. Sloan Foundation, the Participating Institutions, the National Science Foundation, the U.S. Department of Energy, the National Aeronautics and Space Administration, the Japanese Monbukagakusho, the Max Planck Society, and the Higher Education Funding Council for England. The SDSS Web Site is http://www.sdss.org/.

The SDSS is managed by the Astrophysical Research Consortium for the Participating Institutions. The Participating Institutions are the American Museum of Natural History, Astrophysical Institute Potsdam, University of Basel, University of Cambridge, Case Western Reserve University, University of Chicago, Drexel University, Fermilab, the Institute for Advanced Study, the Japan Participation Group, Johns Hopkins University, the Joint Institute for Nuclear Astrophysics, the Kavli Institute for Particle Astrophysics and Cosmology, the Korean Scientist Group, the Chinese Academy of Sciences (LAMOST), Los Alamos National Laboratory, the Max-Planck-Institute for Astronomy (MPIA), the Max-Planck-Institute for Astrophysics (MPA), New Mexico State University, Ohio State University, University of Pittsburgh, University of Portsmouth, Princeton University, the United States Naval Observatory, and the University of Washington.

The PS1 Surveys have been made possible through contributions of the Institute for Astronomy, the University of Hawaii, the Pan-STARRS Project Office, the Max-Planck Society and its participating institutes, the Max Planck Institute for Astronomy, Heidelberg and the Max Planck Institute for Extraterrestrial Physics, Garching, The Johns Hopkins University, Durham University, the University of Edinburgh, Queen's University Belfast, the Harvard-Smithsonian Center for Astrophysics, and the Las Cumbres Observatory Global Telescope Network, Incorporated, the National Central University of Taiwan, and the National Aeronautics and Space Administration under Grant No. NNX08AR22G issued through the Planetary Science Division of the NASA Science Mission Directorate.

Observations reported here were obtained at the MMT Observatory, a joint facility of the Smithsonian Institution and the University of Arizona.

\bibliography{bib}

\end{document}